\documentclass[journal]{IEEEtran}
\usepackage{graphicx,cite,epsfig,amssymb,amsmath}
\usepackage{epstopdf}

\usepackage{pifont}
\usepackage{stmaryrd}
\usepackage{bbding}
\usepackage{algorithm}
\usepackage{algorithmic}
\usepackage{mathrsfs}
\usepackage{latexsym}
\usepackage{indentfirst}
\usepackage{fmtcount}
\usepackage{cite}
\usepackage{float}
\usepackage{morefloats}

\usepackage{caption}
\usepackage{subcaption}

\usepackage{color}
\usepackage{mathtools}
\usepackage{xfrac}

\usepackage{array,colortbl,multirow,multicol,booktabs,ctable,bigstrut}

\usepackage[
top    = 0.75in,
bottom = 0.75in,
left   = 0.75in,
right  = 0.75in]{geometry}

\begin{document}
%
\title{Parity-Check Polar Coding for 5G and Beyond}
%
%
%

\author{
    \IEEEauthorblockN{Huazi~Zhang, Rong~Li, Jian~Wang, Shengchen~Dai, Gongzheng~Zhang, Ying~Chen, Hejia~Luo, Jun~Wang}\\
    \IEEEauthorblockA{Huawei Technologies Co. Ltd.}\\
    Email: \{zhanghuazi,rongone.li,justin.wangjun\}@huawei.com
}
\maketitle

\begin{abstract}
In this paper, we propose a comprehensive Polar coding solution that integrates reliability calculation, rate matching and parity-check coding. Judging a channel coding design from the industry's viewpoint, there are two primary concerns: (i) low-complexity implementation in application-specific integrated circuit (ASIC), and (ii) superior \& stable performance under a wide range of code lengths and rates. The former provides cost- \& power-efficiency which are vital to any commercial system; the latter ensures flexible and robust services. Our design respects both criteria. It demonstrates better performance than existing schemes in literature, but requires only a fraction of implementation cost. With easily-reproducible code construction for arbitrary code rates and lengths, we are able to report ``1-bit'' fine-granularity simulation results for thousands of cases. The released results can serve as a baseline for future optimization of Polar codes.\footnote{The work was first disclosed in 2016 as a technical contribution \cite{3GPP:PC_87} and accepted by IEEE ICC 2018. Part of the proposed design has been adopted by 3GPP as the Polar coding standards for 5G \cite{3GPP:ChairNote_AH2}.}
\end{abstract}


\IEEEpeerreviewmaketitle

\section{Introduction}\label{section:intro}
\subsection{Background and related works}
Answering the question of ``what will 5G be?'' \cite{5G:what}, the result is clear at least for channel coding. For the enhanced Mobile Broadband (eMBB) service category in 5G, LDPC codes and Polar codes \cite{Polar:Arikan,Polar:Stolte} have been adopted for data channel and control channel, respectively. With state-of-the-art code construction techniques \cite{Polar:GA,Polar:CA_List_Niu} and list decoding algorithm \cite{Polar:List_Tal}, Polar codes demonstrate competitive performance under short information block length (K$<$1000), whereas the block error rate (BLER) gain over LDPC and Turbo codes is up to 1dB. Such advantages make Polar codes the most suitable candidate for the control channel, where the payload size is relatively small.

Polar code construction refers to determining the sets of information/frozen bits given certain information block length $K$ and code length $N$. According to \cite{Polar:Arikan,Polar:Stolte}, the most reliable synthesized sub-channels should be selected as information set to obtain the best performance under successive cancellation (SC) decoding. Gaussian approximation (GA) \cite{Polar:GA} is an efficient way to compute the ``reliability" under AWGN channel.

While the performance of an SC decoder is worse than LDPC and Turbo, CRC-aided Polar (CA-Polar) codes \cite{Polar:CA_List_Niu} demonstrate significantly better performance under successive cancellation list (SCL) decoding \cite{Polar:List_Tal}. The reason lies in that the native code distance of Polar codes is relatively poor compared to Reed-Muller codes and many other modern codes. Without CRC bits, an SCL decoder relies solely on path metrics to select from the surviving paths. Thus, codes with poor distance spectrum cannot perform well. In contrast, CA-Polar relies on both path metric and CRC bits to pick the final path, therefore does not suffer from the performance bottleneck incurred by poor code distance.

Although SCL significantly improves the performance of Polar codes, the optimal code construction under list decoding remains an open problem. Beyond CA-Polar, several attempts \cite{Polar:Subcode,Polar:PCC} have been made to design better Polar codes for SCL decoder. A more general form of outer codes, coined as parity-check coding, was introduced to provide additional performance gain as well as flexibility. Polar subcodes \cite{Polar:Subcode} allow some ``dynamic'' frozen bits to be information-bits-dependent. Extended BCH codewords were leveraged to establish parity-check functions such that the constructed codes has guaranteed minimum distance, which is always better than the original Polar codes with the same code length and code rate. Later, a heuristic parity-check construction was introduced in \cite{Polar:PCC}, which also shows evident performance gain over CA-Polar codes. These methods opened a door for better Polar construction with parity check bits.

\subsection{Motivation and our contributions}\label{section:Contribution}
Despite the rich literature on Polar code construction, we found that none of them can be directly applied to a commercial network such as 5G. The reasons are below:
\begin{itemize}
  \item \emph{Implementation complexity:} existing code construction schemes, including rate matching \cite{Polar:QUP,Polar:WangLiuShorten} and parity-check coding \cite{Polar:Subcode,Polar:PCC}, rely heavily on density evolution (DE) (or its simplification GA \cite{Polar:GA}) to acquire sub-channel reliability. These operations (e.g., float-point computations of $\phi(x),\phi^{-1}(x)$ and sorting) are suitable for software simulations but are not hardware-friendly. They either incur large encoding/decoding latency if calculated online, or occupy much memory if calculated offline and pre-stored in ASIC.
  \item \emph{Incomplete solution:} existing parity-check coding schemes are not co-designed with a practical rate-matching scheme. The construction in \cite{Polar:Subcode} is based on $2^m$-length eBCH codewords, and the corresponding generalization to arbitrary code lengths is unknown. The heuristic method in \cite{Polar:PCC} recursively establishes parity-check functions based on GA-acquired reliability. Similarly, a rate-compatible design is not available.
  \item \emph{Lack of fine-granularity evaluation:} existing works \cite{Polar:Subcode,Polar:PCC,Polar:puncture,Polar:QUP,Polar:WangLiuShorten} often draw conclusions from a few special cases (e.g., $N=1024,K=512$). We find it quite common that a scheme that excels in certain cases may perform poorly in other cases, thus their conclusions may not hold for the general cases. To fully evaluate a scheme before large-scale implementation, fine-granularity simulations covering various code lengths and rates are necessary.
\end{itemize}

To address the above issues, we propose a PC-Polar design that integrates deterministic reliability ordering and rate matching schemes. Based on distance spectrum analysis and error propagation patterns, we propose to select PC bits from sub-channels of low row weights, and establish PC functions through a fixed-length cyclic shift register. The entire solution is hardware-friendly to facilitate ASIC implementation. To our best knowledge, such a comprehensive yet low-complexity solution for Polar construction has not been elaborated in literature. Moreover, we provide fine-granularity simulation results to demonstrate stable \& better performance than existing schemes under thousands of cases. Given the construction details, our design should be reproducible for arbitrary code lengths and rates. Therefore, we hope it serve as a baseline for further optimizations of Polar codes.

\section{Polar codes}\label{section:Polar}
A binary Polar code of mother code length $N=2^n$ can be defined by $\textbf{c}_0^{N-1}=\textbf{u}_0^{K-1}\textbf{G}_\mathcal{I}$, where $\textbf{u}_0^{K-1}$ and $\textbf{c}_0^{N-1}$ are message and codeword vectors, respectively, and $\textbf{G}_\mathcal{I}$ is the generator matrix. To construct a $(N,K)$ Polar code, $\textbf{G}_\mathcal{I}$ is obtained by taking rows with indices $i \in \mathcal{I}$ from the $N \times N$ matrix $\textbf{G} = \textbf{F}^{\otimes n}$, where $\mathcal{I}$ is the information sub-channel indices, $\textbf{F} = \begin{bmatrix}1 & 0 \\ 1 & 1\end{bmatrix}$ is the kernel and $^\otimes$ denotes Kronecker power.

\subsection{Reliability ordering}\label{section:PW}
One key step of Polar code construction is determining the information set $\mathcal{I}$. According to Arikan \cite{Polar:Arikan}, the reliability metric is \emph{channel dependent}. Applying this principle, density evolution (DE) (or its simplification Gaussian approximation (GA) \cite{Polar:GA}) calculates the reliability of each synthesized sub-channel based on channel state information (CSI), which can be signal-to-noise-ratio (SNR) or erasure probability. The $K$ most reliable sub-channels are selected as $\mathcal{I}$. In the absence of assistant bits such as CRC or PC bits, the rest $N-K$ sub-channels are selected as the frozen set, denoted by $\mathcal{F}$.

Regarding ASIC implementation, the channel-dependent GA/DE method is infeasible due to (i) float-point computations of complicated functions such as $\phi(x)$, $\phi^{-1}(x)$ and sorting, and (ii) imperfect CSI estimation.

Alternatively, we propose a \emph{channel-independent} Polarization Weight (PW) method as follows. Given a sub-channel index $i$ and its binary expansion $\mathbf{B} = \left(b_{n-1},\cdots,b_1,b_0\right)$, its PW value is defined as
\begin{equation}\label{equ:PW}
W_i \triangleq \sum\limits_{j=0}^{n-1} b_j \beta^j,
\end{equation}
where $\beta$ is empirically chosen to be $2^{\frac 1 4}$ \cite{Polar:PW}. A higher PW value indicates a higher reliability.

A reliability ordered sequence $Q_0^{N-1}$ is obtained \emph{offline} through Algorithm~\ref{alg:PW}, and pre-stored in ASIC such that no on-the-fly calculation is required.
\begin{algorithm}
\begin{algorithmic}
\STATE 1) Calculate $W_i, \forall i \in [0,1,\cdots,N-1]$ according to \eqref{equ:PW}.
\STATE 2) Sort $W_0^{N-1}$ in ascending order.
\STATE 3) Obtain a reliability ordered sequence $Q_0^{N-1}$, such that $W_{Q_0} \leq W_{Q_1} \leq W_{Q_2} \leq \cdots W_{Q_{N-1}}$.
\end{algorithmic}
\caption{Polarization Weight (PW) algorithm}
\label{alg:PW}
\end{algorithm}

\emph{Remark:} Although sub-channel reliability is channel-dependent, \emph{their relative ordering is almost channel-independent under a practical working point} (e.g., BLER within $10^{-4} \sim 10^{-1}$). The simple and closed-form PW formula in \eqref{equ:PW} well approximates this ordering by capturing the recursive polarization process of Polar codes. It generates an information set $\mathcal{I}$ very similar to that generated by GA/DE methods, but requires only a fraction of implementation cost.

\subsection{Rate matching}\label{section:RateMatching}
Rate matching bears much practical importance because, in a commercial system, the allocated channel resource may not have exactly $N = 2^n$ bits. To support an arbitrary code length of $M$, puncturing \cite{Polar:puncture,Polar:QUP} and shortening \cite{Polar:WangLiuShorten} are performed. A well-designed rate matching scheme should bring minimum performance loss with respect to its mother code of length $N$.

For puncturing, $N-M$ bits are not transmitted and deemed \emph{unknown} at the decoder, whereas the log-likelihood ratio (LLR) input of the corresponding punctured position is set to zeros. For shortening, $N-M$ bits are not transmitted and deemed \emph{known} at the decoder, whereas the LLR input of the corresponding shortened position is set to infinite large (see \cite{Polar:puncture,Polar:QUP,Polar:WangLiuShorten} for details).

Quasi-uniform-puncturing (QUP) \cite{Polar:QUP} sequentially punctures the first $N-M$ coded bits, i.e., $\textbf{c}_0^{N-M-1}=[c_0,c_1,\cdots,c_{N-M-1}]$ from the mother codeword $\textbf{c}_0^{N-1}$, and re-calculates the reliability of all $N$ sub-channels using GA. Since the selection of information set $\mathcal I$ fully adapts to the punctured pattern via GA, the method yields good and stable performance under a wide range of code lengths and code rates. The Wang-Liu shortening \cite{Polar:WangLiuShorten} method defines a set of valid shortening patterns based on the Polar kernel, and yields superior performance at higher coding rates. However, both schemes \cite{Polar:QUP,Polar:WangLiuShorten} inherit the same implementation issues from GA, that is, online reliability re-calculations and imperfect CSI estimation.

Similar to \cite{Polar:QUP,Polar:WangLiuShorten}, other existing rate matching schemes rely heavily on re-calculations of sub-channel reliability via GA/DE, since their reliability ordering changes greatly over different punctured/shortened patterns. To implement such schemes, one has to either perform online GA/DE, or pre-store all the $N$-length reliability ordered sequences for each code length $M$. Unfortunately, neither is feasible for ASIC implementation due to complexity/latency and memory constraints.

Our scheme takes the opposite way, i.e., defining a rate matching sequence that, no matter how many bits are punctured/shortened, the pre-defined reliability order (e.g., PW order) is maximally preserved. In this way, only one reliability ordered sequence and another rate matching sequence are required, both of which are of length $N$. 
Furthermore, no online calculation is required. Since the reliability ordered sequence becomes rate-matching independent, inevitable performance loss is incurred. However, the tradeoff is worthwhile given the significant complexity reduction.

The proposed rate matching scheme is described below.
\begin{enumerate}
\item Generate a rate matching pattern $\mathcal R$.
    \begin{itemize}
    \item For shortening, the shortened pattern is defined in Algorithm~\ref{alg:BIV}.
    \item For puncturing, a blockwise-sequential punctured pattern is defined in \cite{3GPP:MTK_puncture}.
    \end{itemize}
\item Select $K$ most reliable sub-channels as $\mathcal I$ according to PW, while skipping the indices in $\mathcal{R}$.
\end{enumerate}

\begin{algorithm}
\begin{algorithmic}
\STATE 1) Define a bit-reversed sequence $T_0^{N-1} = [BR(N-1), BR(N-2), \ \cdots, BR(1), BR(0)]$, where $BR(i)$ denotes the bit-reversed version of $i$. That is, if $i$'s binary expansion is $\left(b_{n-1},\cdots,b_1,b_0\right)$, then $BR(i)$'s binary expansion is $\left(b_0,b_1,\cdots,b_{n-1}\right)$.
\STATE 2) Generate the rate matching pattern $\mathcal R = [T_0,T_1,\cdots,T_{N-M-1}]$, and shorten the corresponding indices in codeword. The transmitted codeword bits are $\hat{c}_0^{M-1} = \left\{c_i \in c_0^{N-1} | i \notin \mathcal{R}\right\}$.
\STATE 3) Freeze the associated sub-channels: $\mathcal R \to \mathcal F$.
\end{algorithmic}
\caption{Bit-Reversed shortening (BRS) algorithm}
\label{alg:BIV}
\end{algorithm}

As mentioned, the rate matching scheme only requires to pre-store $T_0^{N-1}$ (in addition to $Q_0^{N-1}$), thus is hardware friendly. In fact, even $T_0^{N-1}$ can be online generated with simple procedures: switch between big endian and little endian while reading $[N-1,N-2,\cdots,1,0]$, which requires almost no computation overhead.

\section{Parity-check coding}
As mentioned in Section~\ref{section:intro}, CA-Polar improves the performance under list decoding with better distance spectrum. But it has two major limitations. First, CRC bits are essentially independent from the Polar kernel, thus leaves no room for joint optimization. Second, they are appended at the end, thus cannot assist decoding during intermediate decoding stages.

Parity-check bits have the advantage of improving path selection during intermediate decoding stages. Existing parity-check designs are Polar-specific by considering either the Polar kernel \cite{Polar:Subcode}, or its SC decoding process \cite{Polar:PCC}. However, they require high complexity to construct and store the PC functions. Specifically, \cite{Polar:Subcode} requires to perform Gaussian elimination on the parity-check matrix, which has $O(N^3)$ complexity, and \cite{Polar:PCC} requires a recursive algorithm to establish the PC functions. These operations cannot be pipelined for hardware acceleration. Moreover, the PC functions are irregular and do not support compact representation with a few parameters. To implement, a set of bit positions have to be pre-stored to specify each PC function. For example, if a PC function is $u_i+u_j+\cdots+u_k=0$, then the indices $[i,j,\cdots,k]$ are stored, which incurs excessive memory cost especially when the number of PC bits and functions is large.

We address the above problems with a complete solution that integrates our reliability metric in Section~\ref{section:PW} and rate matching scheme in Section~\ref{section:RateMatching}. Our solution is guided by Polar-specific distance spectrum analysis and observations from bit error propagation patterns. The constructed PC functions require only one parameter to represent, and very simple hardware to implement.

\subsection{PC bit positions}
\subsubsection{Distance spectrum analysis}
A distance spectrum analysis of Polar codes can help to select PC bit positions. In an SCL decoder, a path is defined by a binary vector $\mathbf{u}_0^{i-1} = (u_0,u_1,\cdots,u_{i-1}) \in \{0,1\}^i$. At the $i$-th decoding stage, what an SC decoder actually does is deciding whether the received vector is more likely to be from the subset of codewords with $u_i=0$, or the subset of codewords with $u_i=1$.

The former subset is called a \emph{``zero'' coset} and the latter subset is called a \emph{``one'' coset}, respectively defined as
\begin{gather*}
C\left(\mathbf{u}_0^{i-1} | 0\right) = \sum\limits_{j=0}^{i-1}u_j \mathbf{g}_j + span\left(\{\mathbf{g}_{i+1},\cdots,\mathbf{g}_{N-1}\}\right),\\
C\left(\mathbf{u}_0^{i-1} | 1\right) = \mathbf{g}_i + C\left(\mathbf{u}_0^{i-1} | 0\right),
\end{gather*}
where $\mathbf{g}_j$ is the $j$-th row of $\mathbf{G}$, and $C\left(\mathbf{u}_0^{i-1} | 0\right)$ denotes all codewords corresponding to path $\mathbf{u}_0^{i-1}$ and $u_i=0$.

For example, the \emph{``zero'' coset} $C\left(\mathbf{u}_0^{i-1} | 0\right)$ and the \emph{``one'' coset} $C\left(\mathbf{u}_0^{i-1} | 1\right)$ with the same prefix of path $\mathbf{u}_0^{i-1}$ has difference only at $u_i$. The distance spectrum between these two cosets is denoted by $\mathbf{S}^i=\{S_w^i\}, w\in \{0,1,\cdots,N\}$, where
\begin{equation}
S_w^i = \left|\left\{\mathbf{y} \in C\left(\mathbf{0}_0^{i-1} | 1\right), wt(\mathbf{y})=w\right\}\right|,
\end{equation}
where $C\left(\mathbf{0}_0^{i-1} | 1\right)$ denotes all codewords corresponding to path with all ``0'' decoded bits except ``1'' for $u_i$, and $wt(\mathbf{y})$ is the weight (number of non-zero elements) of $\mathbf{y}$.
By the definition of $\mathbf{S}^i$, it is straightforward to see that the minimum distance between the two cosets is
\begin{equation}
\min\left(wt\left(C\left(\mathbf{0}_0^{i-1} | 1\right)\right)\right)=wt(\mathbf{g}_i).
\end{equation}

The concept of cosets naturally extends to an SCL decoder. It is observed that the path metric is closely related to the minimum distance and distance spectrum. To avoid discarding the true path at the $i$-th stage, the path metrics of incorrect paths should receive more penalty than the true path. This can be achieved by letting the cosets induced by different paths to be ``as far as possible'' so that the true path is ``as distinguishable as possible'', especially for paths with differences over only a few bits. In an SCL decoder, ``a larger distance'' between cosets means ``a larger penalty'' on the path metric.

If the $i$-th bit does not involve in any PC function, then the minimum distance between cosets are incurred by the bit positions with minimum row weight (i.e., $wt(\mathbf{g}_i)=w_{min}$) among the unfrozen bits. By selecting these bit positions as PC bits and setting their values using linear combinations of preceding information bits, the path metrics of different paths can be made ``more distinguishable'' and the SCL decoding performance can be improved.

\subsubsection{Tradeoff between reliability and code distance}
As explained, the PC positions should be selected from the unfrozen sub-channel indices with minimum or lower row-weights. However, the number of low-weight positions may be quite large depending on $(N,K)$. It is obviously unwise to select all of them as PC bits. Consider the extreme case where all the low weight positions are selected as frozen bits (can be viewed as a PC bit with PC function $u_i=0$), the remaining information set $\mathcal I$ would be those with the highest row weights and the resulting code construction becomes similar to Reed-Meed codes. Although the distance spectrum of Reed-Muller codes is far better than Polar codes, its BLER performance under SC decoding is poor.

An SCL decoder with practical list sizes (e.g., $L=8$) lies somewhere between an SC decoder and a Maximum Likelihood (ML) decoder. As a result, a good PC-Polar construction should respect both reliability and code distance. In the context of PC-Polar, the corresponding design principle is to pre-select \emph{just enough} PC bits from the most reliable bit positions (those otherwise would be selected as information set $\mathcal I$), such that \emph{the reliability of the remaining information sub-channels are not sacrificed too much}. Note that the unreliable bit positions (those otherwise would be selected as frozen set $\mathcal F$) can be subsequently selected as additional PC bits, which will not sacrifice the reliability of $\mathcal I$.

To summarize, the design principles are:
\begin{itemize}
\item Select the bit positions with \emph{minimum row weights} among the non-frozen bit set as PC bits.
\item Pre-select a \emph{proper number} of PC bits from the reliable bit positions.
\end{itemize}

In practice, easy-to-implement rules must be defined to determine the order for pre-selecting the PC bits. Since the PC functions must be forward-only to be consistent with any SC-based decoder, the last sub-channel index in a PC function always becomes a PC bit. To let the PC functions cover as many information bits as possible. An intuitive way is to select PC bits by descending reliability order\footnote{Since information set $\mathcal I$ is also selected by descending reliability order, the same hardware module can be reused for pre-selecting PC bits.}, such that if an incorrect path passes the parity check, a larger penalty is imposed on its path metric. Specifically, we adopt the following steps:
\begin{enumerate}
  \item Select PC bits from the unfrozen bit positions with the least row weight ($w_{min}$) by descending reliability order.
  \item If there is insufficient unfrozen bit positions with row weight $w_{min}$, continue to select those with row weight $2\times w_{min}$ by descending reliability order.
\end{enumerate}

\subsection{PC functions}
As discussed, the PC bit values should be set to a \emph{linear combination} of some preceding information bits, such that code distance spectrum is improved.

Take $N=16$ for example, if $u_{10}$ is selected as a PC bit, a good PC function would be $u_5 + u_{10} = 0$. Their corresponding row vectors are
\begin{gather*}
\textbf{g}_5 = [1 1 0 0 1 1 0 0 0 0 0 0 0 0 0 0],\\
\textbf{g}_{10} = [1 0 1 0 0 0 0 0 1 0 1 0 0 0 0 0].
\end{gather*}

Observe that $wt(\mathbf{g}_5) = wt(\mathbf{g}_{10}) = 4$. If $u_5$ was an information bit and $u_{10}$ was a frozen bit, the minimum code weight would be at most 4, corresponding to $\textbf{g}_5$ as the lowest-weight non-zero codeword. Now that we change $u_{10}$ into a PC bit, and impose $u_5 = u_{10}$ as a PC function, the combined codeword becomes
\begin{equation*}
\textbf{g}_{5} + \textbf{g}_{10} = [0 1 1 0 1 1 0 0 1 0 1 0 0 0 0 0],
\end{equation*}
which has a higher weight of 6.

For longer codes, it becomes non-trivial to find all the PC functions that improves the minimum code distance. Even if such a method exists, the construction complexity may not be affordable in ASIC. Therefore, we resort to a hardware-friendly way to establish effective PC functions.

From the decoding perspective, $u_5 + u_{10} = 0$ is an effective PC function since it includes sub-channels with relatively independent bit errors. For example, if the $i$-th and $j$-th sub-channels belong to the same PC function, and a bit error in the $i$-th sub-channel leads to another bit error in the $j$-th sub-channel, this $(u_i,u_j)$ error pattern would not be detected by a PC bit. Although bit error propagation is inevitable with SC-based Polar decoding, we should exploit its bit error patterns to mitigate its adversary effect on PC functions.

By Monte-Carlo simulation of a length-16 Polar block, we found that among the $2^{16}-1=65535$ possible error patterns, only 16 of them are dominant and take up around $80\%$ of the total error events. Besides the single error pattern $\mathbf{e}_1 = u_0$, the frequent error propagation patterns are

{\begin{small}
\begin{gather*}
\mathbf{e}_2 = u_0,u_1 \quad \mathbf{e}_3 = u_0,u_2 \quad \mathbf{e}_4 = u_0,u_4 \quad \mathbf{e}_5 = u_0,u_8 \\
\mathbf{e}_6 = u_0,u_1,u_2,u_3 \quad \mathbf{e}_7 = u_0,u_1,u_4,u_5\\
\mathbf{e}_8 = u_0,u_1,u_8,u_9 \quad \mathbf{e}_9 = u_0,u_2,u_4,u_6\\
\mathbf{e}_{10} = u_0,u_2,u_8,u_{10} \quad \mathbf{e}_{11} = u_0,u_4,u_8,u_{12}\\
\mathbf{e}_{12} = u_0,u_1,u_2,u_3,u_4,u_5,u_6,u_7\\
\mathbf{e}_{13} = u_0,u_1,u_2,u_3,u_8,u_9,u_{10},u_{11}\\
\mathbf{e}_{14} = u_0,u_1,u_4,u_5,u_8,u_9,u_{12},u_{13}\\
\mathbf{e}_{15} = u_0,u_2,u_4,u_6,u_8,u_{10},u_{12},u_{14}\\
\mathbf{e}_{16} = u_0,u_1,u_2,\cdots,u_{14},u_{15}
\end{gather*}
\end{small}}
Observe that the most frequent error patterns are between every 1, 2, 4, 8 bit positions. This is due to the power-of-2 recursive structure in Polar kernel. Intuitively, we should avoid setting up PC functions over bit positions with power-of-2 spacings. In contrast, we found that bit errors propagate less frequently between every 5 bit positions.

An effective yet implementable way is to set up PC functions over bit positions with fixed $p$-sized spacing, where $p$ can be set to 5 for all cases. It can be easily implemented by a $p$-length cyclic shift register (CSR). The PC pre-coding function, denoted by $\mathcal{F}_{PC}: u_0^{K-1} \to \hat{u}_0^{N-1}$, is described by Algorithm~\ref{alg:PC_precoding}.
\begin{algorithm}
\begin{algorithmic}
\STATE \textbf{Initialization:} $y[0],\cdots,y[p-1] = 0$, $k=0$
\FOR {$i$ in $[0,1,\cdots,N-1]$}
\STATE Cyclic shift the register
\STATE If the $i \in \mathcal{I}$, then set $\hat{u}_i=u_k$, update CSR by $y[0] = u_k \bigoplus y[0]$, and count $k=k+1$
\STATE If the $i \in \mathcal{P}$, then set $\hat{u}_i=y[0]$
\STATE If the $i \in \mathcal{F}$, then set $\hat{u}_i=0$
\ENDFOR
\end{algorithmic}
\caption{PC pre-coding algorithm}
\label{alg:PC_precoding}
\end{algorithm}

A PC decoder reuses the same algorithm, in which $u_k$ is the decoded value of an information bit, and the expected PC bit value is the first register state $y[0]$ for $i\in \mathcal P$. All paths with an unexpected PC bit value are pruned.

\begin{figure}
\centering
    \includegraphics[width= 0.40\textwidth]{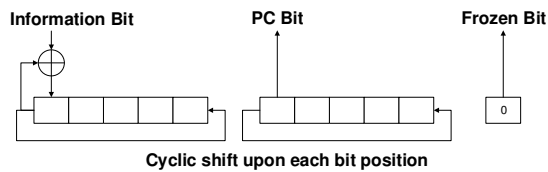}
    \caption{Cycle shift register implementation of PC functions.}
    \label{fig:register}
\end{figure}

The equivalent CSR operation is shown in Figure~\ref{fig:register}. It has the following advantages:
\begin{itemize}
  \item The PC function has only one parameter $p$. No need to feed the constructor with every individual PC function.
  \item The complexity does not grow with the number of PC bits or PC functions. All of them can be implemented by a single set of CSR.
  \item The encoder and decoder can share the same CSR to further save chip area.
\end{itemize}

Note that more sophisticated multiple feedback CSR can also be adopted, which is defined by a polynomial. However, the implementation in Figure~\ref{fig:register} with $p=5$ is the simplest while preserves the best performance.

\subsection{Code construction algorithm}
\begin{figure*}
\centering
    \includegraphics[width= 0.75\textwidth]{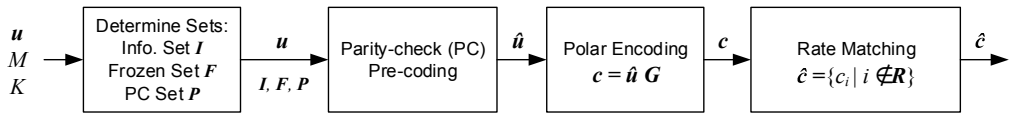}
    \caption{Parity-check Polar code construction flow.}
    \label{fig:construction_flow}
\end{figure*}

A full code construction flow is depicted in Figure~\ref{fig:construction_flow}, in which the PC pre-coding module is described in Algorithm~\ref{alg:PC_precoding} and the information/frozen/PC set generation module is detailed in Algorithm~\ref{alg:bit_set}. The rate matching pattern $\mathcal R$ is obtained according to Algorithm~\ref{alg:BIV}.
\begin{algorithm}
\begin{algorithmic}
\STATE \textbf{Initialization:} code length $M$, mother code length $N = 2^{\lceil\log_2(M)\rceil}$, information length $K$
\STATE \textbf{Generate reliability ordered sequence $Q_0^{N-1}$:} use the PW method (Algorithm~\ref{alg:PW}).
\STATE \textbf{Generate rate matching pattern $\mathcal R$:} use the BRS method (Algorithm~\ref{alg:BIV}).
\STATE \textbf{Determine parameters $(w_{min},f_1,f_2)$:}
\STATE 1) Estimate $f=\log_2N \times \left(\alpha-|\alpha \times (K/M-1/2)|^2\right)$ as an estimated number of pre-selected PC bits.
\STATE 2) Determine $w_{min}$ as the smallest row weight within the $K+f$ most reliable sub-channels (excluding the sub-channels with indices in $\mathcal R$), and count the number of such sub-channels as $n_{w_{min}}$.
\STATE 3) Pre-select $f_1$ and $f_2$ PC bits with row weight $w_{min}$ and $2\times w_{min}$, respectively, according to descending reliability order and skipping the sub-channels with indices in $\mathcal R$.
\STATE \quad If $f \leq n_{w_{min}}$, then $f_1 = f, f_2 = 0$;
\STATE \quad If $f > n_{w_{min}}$, then $f_1 = n_{w_{min}}, f_2 = 3/4(f-n_{w_{min}})$.
\STATE \textbf{Generate $\mathcal I$, $\mathcal F$ and $\mathcal P$:}
\STATE 4) Generate $\mathcal I$ by selecting $K$ information bits according to descending reliability order, while skipping the indices in $\mathcal R$ and the pre-selected $f_1 + f_2$ PC bits.
\STATE 5) Generate $\mathcal P$ by selecting all the remaining sub-channels except those in $\mathcal R$, that is, $\mathcal P = \{i:i\notin \mathcal{I},i\notin \mathcal{R}\}$.
\STATE 6) Generate $\mathcal F$ by selecting the bit positions in $\mathcal R$.
\end{algorithmic}
\caption{Information/Frozen/PC bit set selection}
\label{alg:bit_set}
\end{algorithm}

Some clarifications to Algorithm~\ref{alg:bit_set} are as follows. Step 1$\sim$3 can be performed once offline for faster construction, and the parameter tuple $(w_{min},f_1,f_2)$ can be pre-stored. There are two types of PC bits, i.e., the ``reliable'' PC bits pre-selected in Step 3 and the ``unreliable'' PC bits\footnote{The PC bits before the first information bit are equivalent to frozen bits.} additionally selected in Step 5. The rough number of pre-selected PC bits $f$ is determined based on our observation that codes with rate near $1/2$ require more PC bits than higher and lower rates. In addition, $f$ is upper bounded by $(M-K)/2$. The coefficient $\alpha$ is used to control the number of pre-selected PC bits. The larger $\alpha$ is, the more PC bits are pre-selected\footnote{Note that other ways to control the number of pre-selected PC bits are allowed as long as they produce good performance.}. Typically, a smaller $\alpha$ can be used for an SCL decoder with smaller list sizes, and a larger $\alpha$ can be used for an SCL decoder with larger list sizes and better performance at higher SNR region. To facilitate reproducible research, we set $\alpha=1$ in all our simulations for a balanced performance under an SCL decoder with a practical list size $L=8$.


\section{Simulation results}
To validate the proposed PC-Polar design, we not only compare with existing Polar coding schemes, but also provide ``1-bit'' fine-granularity simulation results covering a wide range of code lengths and rates. A parity-check (PC) SCL decoder is used for PC-Polar codes. It is similar to an SCL decoder except that it only keeps paths that satisfy PC functions during intermediate decoding stages. The CRC polynomials we use for CA-Polar are $D^8 +D^7 +D^6 +D^3 +D^2 +D +1$ (CRC8) and $ D^{16} +D^{12} +D^{11} +D^9 +D^8 +D^5 +D^3 +D +1$ (CRC16).
\subsection{PC bit gain}
\subsubsection{Comparison with CA-Polar}
\begin{figure}
\centering
    \includegraphics[width= 0.45\textwidth]{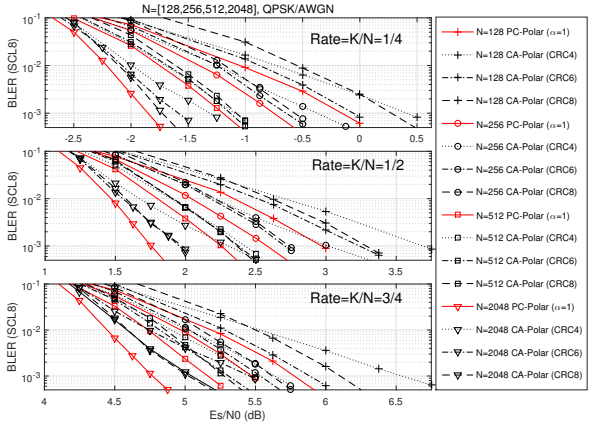}
    \caption{PC-Polar's gain over CA-Polar with different CRC lengths at various code rates under SCL decoder with $L=8$.}
    \label{fig:PCvsCA_ByRate}
\end{figure}

In Figure~\ref{fig:PCvsCA_ByRate}, we compare PC-Polar with CA-Polar under various \emph{mother code lengths}. The reliability ordering for PC-Polar is obtained by the hardware-friendly PW method, and that of CA-Polar is obtained by the computation-intensive GA method. The comparison is actually unfair for PC-Polar in terms of performance, since GA is more precise while PW is only an approximation. Nevertheless, we observe that PC-Polar still outperforms CA-Polar at all cases. This is due to both the sufficient gain from PC bits and negligible loss from the PW method.

It is also observed that, as the number of CRC bits increases, the performance of CA-Polar fails to improve after the CRC length reaches 8. The best performance achieved by that of CA-Polar is still worse than that of PC-Polar.

\begin{figure}
\centering
    \includegraphics[width= 0.45\textwidth]{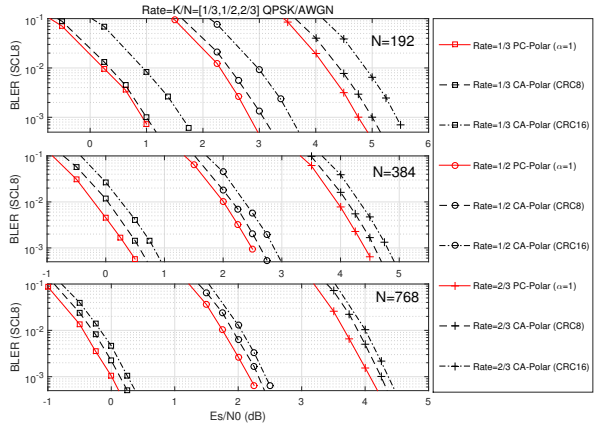}
    \caption{PC-Polar with BRS vs CA-Polar with QUP at various code lengths under SCL decoder with $L=8$.}
    \label{fig:PCvsCA_ByN}
\end{figure}

In Figure~\ref{fig:PCvsCA_ByN}, we simulate the cases with \emph{non-mother code lengths}, where the rate matching methods are BRS for PC-Polar and QUP for CA-Polar, respectively. Again, the construction complexity for former is much lower than the latter, since reliability re-ordering with respect to different rate-matching patterns is not allowed in the BRS method. Similarly, stable PC bit gain is observed. The overall gain is up to 0.8dB compared with CA-Polar (CRC16) and 0.3dB compared with CA-Polar (CRC8).

\subsubsection{Comparison of different parity-check schemes}
We further compare the proposed PC-Polar scheme with existing parity-check schemes such as eBCH-Polar \cite{Polar:Subcode} and PCC \cite{Polar:PCC}. Since both \cite{Polar:Subcode,Polar:PCC} only provided construction procedures and simulation results under mother code lengths, and the associated construction parameters (e.g., the design distance $d$ for eBCH-Polar and the number of check bits $cK$ for PCC) are available only for a few cases, our comparison focuses on these reproducible cases. CA-Polar with 8-bit and 16-bit CRC is also simulated for reference. For CA-Polar, eBCH-Polar and PCC, the GA method is applied to obtain a more precise reliability ordering; for PC-Polar the low-complexity PW method is applied.

\begin{figure}
\centering
    \includegraphics[width= 0.45\textwidth]{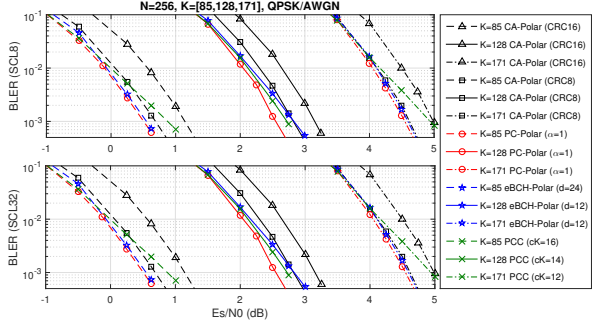}
    \caption{Comparison of existing parity-check-based schemes under $N=256$ and SCL decoder with $L=8,32$.}
    \label{fig:PC_eBCH_PCC_N256}
\end{figure}

\begin{figure}
\centering
    \includegraphics[width= 0.45\textwidth]{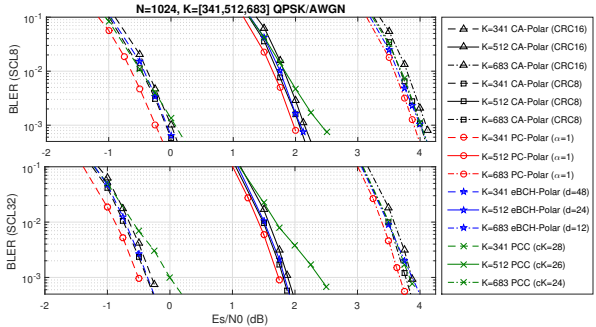}
    \caption{Comparison of existing parity-check-based schemes under $N=1024$ and SCL decoder with $L=8,32$.}
    \label{fig:PC_eBCH_PCC_N1024}
\end{figure}

As shown in Figures~\ref{fig:PC_eBCH_PCC_N256} and \ref{fig:PC_eBCH_PCC_N1024}, all the parity-check based schemes (except for PCC under two cases) have better performance than CA-Polar (CRC16) at the working point of interest, i.e., $BLER=10^{-2}\sim10^{-3}$, which confirms the results reported in \cite{Polar:Subcode,Polar:PCC}. In particular, we found that eBCH-Polar exhibits more stable performance than PCC due to the minimum-distance-guaranteed construction algorithm. PCC also has good performance under most cases, especially at low SNR region.

Among these schemes, PC-Polar demonstrates the best performance in all cases. The gain over CA-Polar with 16-bit and 8-bit CRC is 0.8dB and 0.3dB, respectively. The gain over eBCH-Polar and PCC varies over different cases. In certain cases, PC-Polar has slightly better performance than eBCH-Polar and PCC; while in a few cases, the gain of PC-Polar can reach 0.5dB.

\subsection{Fine-granularity simulations}
As observed in Figure~\ref{fig:PC_eBCH_PCC_N256} and \ref{fig:PC_eBCH_PCC_N1024}, a scheme with excellent performance in one case may have worse performance in other cases. In order to draw more solid conclusion based on more simulation cases, fine-granularity simulation is necessary in the evaluation of channel coding schemes.

Therefore, we conduct ``1-bit'' granularity ($K=8,9,10,\cdots,800$) to cover a wide range of \emph{mother} and \emph{non-mother} code lengths, and typical code rates that are used in control and data channels..

\begin{figure}
\centering
    \includegraphics[width= 0.45\textwidth]{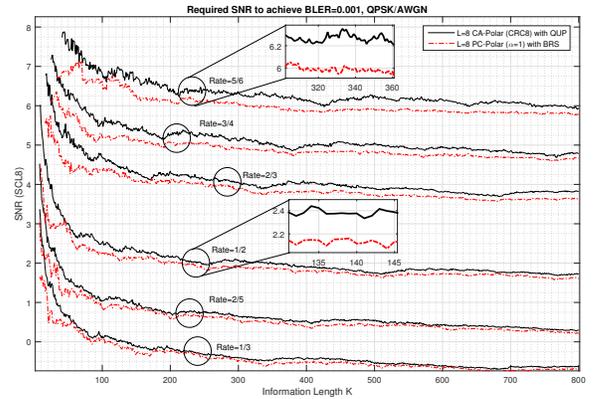}
    \caption{PC-Polar with BRS vs CA-Polar with QUP at $Rate=[1/3,2/5,\cdots,3/4,5/6]$ under SCL decoder with $L=8$.}
    \label{fig:PCvsCA_FineGran_RateMatch}
\end{figure}

In Figure~\ref{fig:PCvsCA_FineGran_RateMatch}, we report the required SNR to achieve $BLER=0.001$ for PC-Polar and CA-Polar in over 4700 cases. The gain ranges from 0.2dB to 1dB. Similar to previous experiments, GA/QUP are applied in CA-Polar and PW/BRS are applied in PC-Polar. For CA-Polar, 8-bit CRC instead of 16-bit CRC is adopted for better performance. Even though, these extensive results clearly show that PC-Polar outperforms CA-Polar in almost all cases. The results demonstrate that PC-Polar has stable \& better performance
than CA-Polar in terms of error correction capability.

\section{Conclusion}
In this work, we propose a novel Polar construction with superior \& stable error correction performance under a wide range of code rates and lengths. As a full solution that integrates hardware-friendly reliability ordering, rate matching and parity-checking methods, our design moves one further step beyond CA-Polar and is implementable for 5G and future networks. Our solution, as detailed in this paper, applies for arbitrary code lengths and rates. Its performance can be reproduced to serve as a baseline for further optimizations.

\ifCLASSOPTIONcaptionsoff
  \newpage
\fi


\begin{thebibliography}{1}
\bibitem{5G:what}
J. Andrews, S. Buzzi, W. Choi, S. Hanly, A. Lozano, A. Soong, J. Zhang, ``What will 5G be?'', \textit{IEEE Journal on Selected Areas in Communications}, vol. 32, no. 6, pp. 1065--1082, Jun. 2014.

\bibitem{Polar:Arikan}
E. Arikan, ``Channel polarization: A method for constructing capacity-achieving codes for symmetric binary-input memoryless channels'', \textit{IEEE Trans. Inf. Theory}, vol. 55, no. 7, pp. 3051--3073, Jul. 2009.

\bibitem{Polar:Stolte}
N. Stolte, ``Recursive codes with the Plotkin-Construction and their Decoding'', Ph.D. dissertation, University of Technology Darmstadt, Germany.

\bibitem{Polar:CA_List_Niu}
K. Niu and K. Chen, ''CRC-aided decoding of polar codes'', \textit{IEEE Communications Letters}, vol. 16, no. 10 pp. 1668--1671, Oct. 2012.

\bibitem{Polar:GA}
P. Trifonov, ``Efficient design and decoding of polar codes'', \textit{IEEE Transactions on Communications} vol. 60, no. 11 pp. 3221--3227, Nov. 2012.

\bibitem{Polar:List_Tal}
I. Tal, and A. Vardy, ``List decoding of polar codes'', \textit{IEEE Trans. Inf. Theory}, vol. 61, no. 5 pp. 2213--2226, May 2015.

\bibitem{Polar:Subcode}
P. Trifonov and V. Miloslavskaya, ``Polar Subcodes'', \textit{IEEE Journal on Selected Areas in Communications}, vol. 34, no. 2, pp. 254--266, Feb. 2016.

\bibitem{Polar:PCC}
T. Wang, D. Qu and T. Jiang, ``Parity-check-concatenated polar codes'', \textit{IEEE Communications Letters}, vol. 20, no. 12 pp. 2342--2345, Dec. 2016.

\bibitem{3GPP:PC_87}
R1-1611254 ``Details of the polar code design'', Huawei, HiSilicon, 3GPP TSG RAN WG1 \#87 Meeting, Reno, USA, Nov. 10th--14th, 2016.


\bibitem{3GPP:ChairNote_AH2}
``Chairman's notes: RAN1'', 3GPP TSG RAN WG1 NR Ad-Hoc Meeting \#2, Qingdao, China, 27th--30th Jun. 2017.

\bibitem{Polar:PW}
X. Liu \emph{et al.}, ``$\beta$-expansion A Theoretical Framework for Fast and Recursive Construction of Polar Codes'' in \emph{Proc IEEE Globecom}, Dec. 2017.

\bibitem{Polar:puncture}
L. Zhang, Z. Zhang, X. Wang, Q. Yu and Y. Chen, ``On the puncturing patterns for punctured polar codes'', in \emph{Proc IEEE ISIT}, pp. 121-125, Jun. 2014.

\bibitem{Polar:QUP}
K. Niu, K. Chen and J. R. Lin, ``Beyond Turbo codes: Rate-compatible punctured Polar codes,'' in \emph{Proc IEEE ICC}, pp. 3423--3427, Jun. 2013.

\bibitem{Polar:WangLiuShorten}
R. Wang and R. Liu, ``A novel puncturing scheme for polar codes,'' \emph{IEEE Communications Letters}, vol. 18, no. 12, pp. 2081--2084, 2014.

\bibitem{3GPP:MTK_puncture}
R1-167533, ``Examination of NR coding candidates for low rate applications'', MediaTek Inc., 3GPP TSG RAN WG1 \#86 Meeting, Gothenburg, Sweden, Aug. 22nd--26th, 2016.

\end{thebibliography}
\end{document}